\documentclass[12pt]{iopart}
\usepackage{graphicx}
\begin{document}

\title[Gravitational-Wave Astronomy]{Gravitational-Wave Astronomy:\\ Observational Results and Their Impact}

\author{Peter S Shawhan}

\address{The University of Maryland, College Park, MD 20742, USA}
\ead{pshawhan@umd.edu}
\begin{abstract}
  The successful construction and operation of highly sensitive
  gravitational-wave detectors is an achievement to be proud of, but
  the detection of actual signals is still around the corner.  Even
  so, null results from recent searches have told us some interesting
  things about the objects that live in our universe, so it can be
  argued that the era of gravitational-wave astronomy has already
  begun.  In this article I review several of these results and
  discuss what we have learned from them.  I then look into the
  not-so-distant future and predict some ways in which the detection
  of gravitational-wave signals will shape our knowledge of
  astrophysics and transform the field.
\end{abstract}

\submitto{\CQG}

\section*{Prologue}

It has been 50 years since Joseph Weber first embarked on a
serious experimental program to try to detect gravitational waves
directly~\cite{weber60}, motivated by the possibility of detecting
signals from sources such as a core-collapse supernova or a binary
neutron star~\cite{weber66}.  The intervening years have seen great
advances in technologies and new techniques for detecting
gravitational waves, from much-improved ``Weber bars''
to highly sensitive broadband interferometers,
Doppler tracking of spacecraft such as Cassini~\cite{CassiniGW},
long-term campaigns to monitor pulse arrival times from stable
pulsars~\cite{PulsarTiming}, and mature plans for long-baseline
interferometer networks in space (namely LISA~\cite{LISA} and
DECIGO~\cite{DECIGO}).
In parallel, after the discovery of the first binary pulsar in
1974~\cite{HulseTaylor}, radio pulse timing campaigns with a number of
short-period binary pulsars have provided compelling ``indirect''
evidence for the existence of gravitational radiation as well as
precise experimental tests of the general theory of
relativity~\cite{DoublePulsar}.
Theoretical work and numerical modeling have provided a
much better understanding
of the likely gravitational-wave (GW) signatures of the original
leading source candidates---core-collapse supernovae~\cite{OttSNGW}
and binary neutron stars~\cite{Blanchet35}---as well as many other
expected or plausible GW sources, including binary systems with
supermassive or stellar-mass black holes,
short-period white dwarf binaries, non-axisymmetric spinning or
perturbed neutron stars, cosmic strings, early-universe processes, and
more.  (General overviews of GW sources may be found in
\cite{CutlerThorne} and \cite{SchutzOverview}, for instance.)  And
here is what our direct searches have yielded so far: {\em Nothing}.

The lack of a directly detected signal is not surprising, based on our
limited knowledge of source populations and on the current sensitivity
levels of the detectors.  Further instrumental improvements are on the
way, including substantial upgrades to the current large
interferometers, proposals to build additional interferometers,
pulse timing measurements of more pulsars for longer time spans with
better precision, and eventually the launch of space missions to open
up the low-frequency window that is certain to be rich with signals.
According to current schedules, we are sure to be detecting signals
and doing GW astronomy around the middle of the coming decade.

However, in this article I will argue that we are {\em already} doing
GW astronomy.  In the next section I will summarize and interpret
several completed searches
for which the lack of a detectable signal
provides some relevant information about the population and/or
astrophysics of plausible sources.  I will then project forward to the
time when signals {\em are} detected and discuss what we may learn
from them, and how they will fundamentally change the field of GW
astronomy.

\section{The impact of null results published so far}

In 1969 Weber claimed that his detectors had produced definitive
evidence for the discovery of gravitational waves~\cite{weber69}, the
first in a series of claims by him that could not be reproduced by
others and were ultimately discredited~\cite{collins}.  Nevertheless,
his attempts inspired an experimental community that continued to
improve the detection technologies, cooling bars to cryogenic
temperatures to minimize thermal noise and exploring other detection
methods~\cite{Thorne300yrs}.

Large cryogenic bar
detectors---ALLEGRO~\cite{ALLEGRO}, EXPLORER~\cite{EXPLORER},
NAUTILUS~\cite{NAUTILUS}, NIOBE~\cite{NIOBE}, and
AURIGA~\cite{AURIGA}---
began operating for extended periods with good sensitivity
in the 1990s.  Analyzing the data from two or more detectors
together substantially reduced the false alarm rate from spurious
signals in the individual detectors (from mechanical vibrations,
cosmic rays, etc.)\ and enabled searches for weaker and less-frequent
transient signals.  This culminated in the 1997 formation of the
International Gravitational Event Collaboration (IGEC)~\cite{IGECoper}.
Around the same time, GW searches with prototype interferometric
detectors ({\it e.g.}~\cite{BNS40m}) gave way to the commissioning and
eventual operation of the full-scale interferometers TAMA\,300~\cite{TAMA},
LIGO~\cite{LIGO}, GEO\,600~\cite{GEO} and Virgo~\cite{Virgo}, which have by
now surpassed the bars in searching for ``high-frequency'' GW signals
(above $\sim10$~Hz).
Also, the past several years have seen advances in using multiple
millisecond pulsars to search for GW signals at
frequencies around $10^{-9}$--$10^{-7}$~Hz, for instance with the
Parkes Pulsar Timing Array project~\cite{PPTA}.

Over the past decade, dozens of papers have been published to report
results from searches for various types of GW signals.  Aside from a
few hints of excess event candidates that were not confirmed, all
searches so far have yielded null results.  From these are derived
upper limits on the rate and/or strength of GW signals reaching the
detectors, or alternatively on the possible population of sources.  In
this section I highlight several of these results which represent
significant steps toward gravitational-wave astronomy.  Many of these signals
are expected to be detectable by the current instruments only if they
originate relatively nearby; therefore we begin by exploring our
cosmic neighborhood.

\subsection{Getting to know our neighbors better}

Our galaxy is thought to contain $10^8$--$10^9$ neutron
stars~\cite{NarayanOstriker}, of which a few thousand have been
detected as radio or X-ray pulsars.  A rapidly spinning neutron star
can emit periodic gravitational waves through a number of
mechanisms, including a static deformation that breaks
axisymmetry~\cite{JKS}, persistent matter oscillations ({\it e.g.\
  r}-modes)~\cite{Andersson}, or free precession~\cite{VDBprecession}.

The Crab pulsar, at a distance of about 2 kpc, is a particularly
interesting neighbor.  With a current spin frequency of $29.7$~Hz and
spin-down rate of $-3.7 \times 10^{-10}$~Hz/s~\cite{CrabMonthly}, its
energy loss rate is estimated to be $4 \times 10^{31}$~W.  This powers
the expansion and electromagnetic luminosity of the Crab Nebula, but
the energy flows in this complex system are difficult to pin down
quantitatively~\cite{CrabRev2008}, leaving open the possibility that a
significant fraction of the energy could be emitted as gravitational
radiation.  Palomba has estimated that the observed braking index of
the pulsar spin-down constrains this fraction to be no more than
40\%~\cite{Palomba}.  The LIGO Scientific Collaboration (LSC)
used data from the first 9
months of LIGO's S5 science run to search for a GW signal from the
Crab pulsar and, finding none, were able to set an upper limit of 8\%
of the total spin-down energy, using X-ray image information to infer the
orientation of the pulsar spin axis and assuming that the GW emission is
phase-locked at twice the radio pulse frequency~\cite{Crab9month}.
Analysis of the full S5 data~\cite{KnownPulsarsS5} has improved this
limit to just 2\%.  This observational result
directly constrains
the properties of the Crab pulsar and the energy balance of the
nebula.

\begin{figure}[bt]
\begin{center}
\includegraphics[width=11cm]{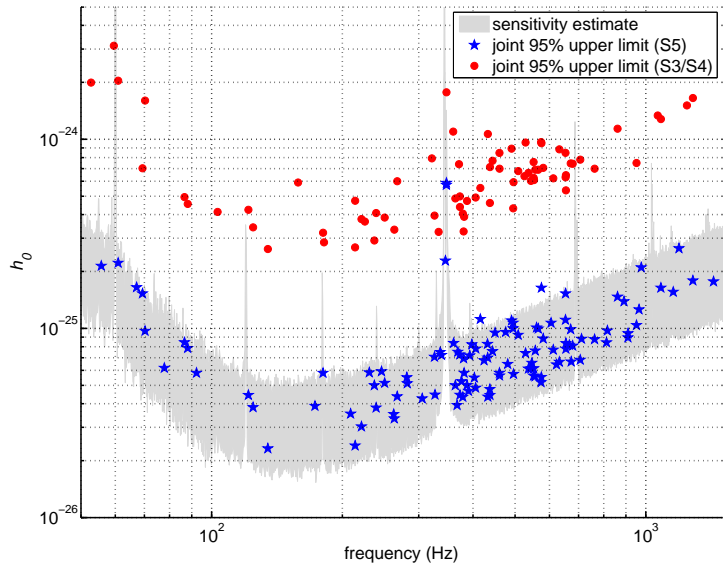}
\caption{Upper limits on GW strain amplitude parameter $h_0$ for known
  pulsars using data from the LIGO/GEO\,600 S3, S4 and S5 science runs,
  taken from~\cite{KnownPulsarsS5} (reproduced by permission
  of the AAS).
  Each symbol represents one pulsar and is plotted at the expected GW
  signal frequency (twice the spin frequency).  The grey band is the
  range expected for S5 given the average instrumental noise level.}
\label{fig:CWResultsPlot}
\end{center}
\end{figure}

The LSC and the Virgo Collaboration (now analyzing data jointly)
have also searched for periodic GWs from all known
pulsars with spin frequencies greater than 20~Hz and sufficiently
precise radio or X-ray pulse timing to allow a fully-coherent search,
again assuming that the GW emission is phase-locked at twice the pulse
frequency.  The analysis~\cite{KnownPulsarsS5} considered 115
radio pulsars (including 71 in binary systems) that were timed during
the LIGO S5 run by the Jodrell Bank Observatory, the Green Bank Telescope,
and/or the Parkes radio telescope, along with the X-ray pulsar
J0537$-$6910 which was monitored by the Rossi X-ray Timing Explorer
(RXTE).  For each pulsar, a 90\% upper limit was placed on the GW
amplitude in terms of the parameter $h_0$, which represents the strain
amplitude that would reach the Earth in the ``plus'' and
``cross'' polarizations {\em if} the pulsar spin axis were oriented
optimally.  Figure~\ref{fig:CWResultsPlot} shows these upper limits,
which are remarkably small numbers in themselves.
The lowest upper limit is $2.3 \times 10^{-26}$, obtained for pulsar
J1603$-$7202.
J0537$-$6910 is the only pulsar besides the
Crab for which the upper limit from this analysis reaches the
spin-down limit, assuming that the moment of inertia is within the
favored range of (1 to 3) $\times 10^{38}$~kg\,m$^2$.

Assuming the neutron stars to be triaxial ellipsoids, these amplitude
limits may be re-cast as limits on the equatorial ellipticity
$\varepsilon$.
Pulsar J2124$-$3358, at
a distance of $\sim$200~pc and GW frequency of 404~Hz, yields the
strictest upper limit, $\varepsilon < 7.0 \times 10^{-8}$.  One may
ask whether the perfect or near-perfect axisymmetry of these neutron
stars is due to the properties of the neutron star material.  It has
long been thought that conventional neutron stars could support an
ellipticity up to a few times $10^{-7}$~\cite{UCB}, but recent work
suggests that the pressure in the crystalline crust suppresses
defects~\cite{HorowitzKadau} and ellipticities of up to $\sim 4 \times
10^{-6}$ are possible in a conventional neutron star.  Most of the
pulsars in the S5 analysis have $\varepsilon$ limits below that level,
meaning that these neutron stars, at least, are closer to axisymmetric
than required by the intrinsic material properties.

\subsection{When the neighbors are disturbed...}

Soft gamma repeaters (SGRs) are believed to be magnetars, {\it i.e.}\
neutron stars with very strong magnetic fields~\cite{DuncanThompson}.
SGRs are observed to emit intense flares of soft gamma rays at
irregular intervals which may be associated with ``starquakes'',
abrupt cracking and rearrangement of the crust and magnetic field.
These events could excite quasinormal modes of the neutron star which
then radiate gravitational waves~\cite{dFPSGRGW}.  The fundamental
mode, at a frequency of around $1.5$ to 3~kHz, is expected to be the
most efficient GW emitter, although other nonradial modes may also
participate.

The LSC have used LIGO data to search for GW bursts
associated with flares of SGRs 1806$-$20 and 1900$+$14, including the
December 2004 giant flare of SGR 1806$-$20~\cite{Horvath}.  A first
analysis~\cite{S5y1SGR} treated the flares individually, setting upper
limits on GW energy as low as a few times $10^{45}$~erg, depending
strongly on the waveform assumed.  The best of these limits (for
signals in the most sensitive range of the instruments, 100--200~Hz)
are within the range of possible GW energy emission during a giant
flare, $10^{45}$ to $10^{49}$~erg, according to modeling by
Ioka~\cite{IokaSGR}.  Unfortunately, the giant flare occurred between
LIGO science runs, and the less-sensitive data available at that time
only yields upper limits on GW energy emission of $5 \times
10^{47}$~erg and above.

A later paper re-analyzed the ``storm'' of SGR 1900+14 flares that
spanned a period of $\sim$30 seconds on 29 March 2006~\cite{SGRstack}.
This analysis ``stacked'' the data around the times of the individual
flares in order to gain sensitivity under the assumption
that many or all of the flares had an associated GW burst at a common
relative time offset.  Two scenarios were considered to choose the
relative weighting of the flares: one in which the GW burst energy is
assumed to be proportional to the electromagnetic fluence of each
flare, and the other in which all large flares are assumed to have
more-or-less equal GW burst energy.  This analysis yielded per-burst
energy upper limits as low as $2 \times 10^{45}$~erg, an order of
magnitude lower than the limits set for this storm by the earlier
single-burst analysis.

These searches are just beginning to address the few existing models
of GW emission by SGRs, and are motivating new modeling of SGRs and
their disturbances.  Stronger constraints (if not a detection) will be
obtained when another giant flare occurs while the GW detector network
is operating with good sensitivity, and/or from searches using
ordinary flares from closer SGRs such as the recently discovered SGRs
J0501$+$4516~\cite{Rea0501,Aptekar0501} and
J0418$+$5729~\cite{vdH0418}, which may both be less than 2~kpc away.

\subsection{Listening for invisible neighbors}

As noted previously, only a small fraction of the hundreds of millions
of neutron stars in our Galaxy are visible to us in radio waves,
X-rays or gamma rays.  It is quite plausible that one or more nearby,
unseen neutron stars have a large enough asymmetry and spin rate to be
emitting periodic gravitational waves at a detectable level.  A
general argument, originated by Blandford and extended in a 2007 paper
by the LIGO Scientific Collaboration~\cite{S2Fstat}, starts with the
(very optimistic) assumptions that all neutron stars are born with a
high spin rate and spin down due to GW emission alone, and concludes
that the strongest signal that we can expect (in an average sense) to
receive on Earth would have $h_0 \simeq 4 \times 10^{-24}$,
independent of frequency and $\varepsilon$.  Knispel and Allen have
greatly refined this analysis~\cite{KnispelAllen}, replacing Blandford's simple
assumptions about the neutron star population with a detailed
simulation of the birth, initial kick and subsequent motion of neutron
stars in the Galaxy.  For a nominal ellipticity of
$10^{-6}$, they find that the maximum expected GW signal amplitude
(with the same optimistic assumptions about neutron stars spinning
down due to GW emission) is around $10^{-24}$ over the frequency range
100--1000~Hz.

The LSC have published two all-sky searches for periodic GW
signals using data from the early part of the S5 run, one using the
semi-coherent ``PowerFlux'' method~\cite{EarlyS5PowerFlux} and the
other using the substantial computing power provided by the
Einstein@Home project to carry out longer coherent integrations on a
smaller data set~\cite{EarlyS5EatH}.  These searches had comparable
sensitivities, both slightly surpassing the Knispel and Allen model
expectations (with $\varepsilon=10^{-6}$) for pulsars with favorable
orientations and GW signal frequencies in the vicinity of 200~Hz.
Thus, periodic GW searches may be on the verge of detecting unseen
neutron stars, or at least constraining models for the population of
such objects in our Galaxy.

\subsection{Listening to the galaxy next door}

On 1 February 2007, an extremely intense gamma-ray burst was detected
by detectors on the Konus-Wind, INTEGRAL, MESSENGER, and
Swift satellites.  The initial position error box from the relative
arrival times of the bursts~\cite{GRB070201loc}
intersected the spiral arms of M31 (the
Andromeda galaxy), raising the intriguing possibility that it
originated in that galaxy, only $\sim$770~kpc away.  Furthermore, the
leading model for most short-hard GRBs such as this one is a binary
merger involving at least one neutron star~\cite{NakarSHGRB}.  Such an
event would also emit strong gravitational waves.

At the time of the GRB, the two detectors at the LIGO Hanford
Observatory were collecting science-mode data, while the other large
interferometers were not.  The LSC searched this data for
both an inspiral signal leading up to the merger and for an arbitrary
GW burst associated with the merger itself~\cite{GRB070201search}.  No
plausible signal was found, and from the absence of a detectable
inspiral signal at that range, a compact binary merger in M31 was
ruled out with $>99\%$ confidence.  The LIGO null result, along with a
refined position estimate for the GRB~\cite{Mazets070201}, helped to
solidify the case that this was most likely an SGR giant flare event
in M31~\cite{Ofek070201}.

\subsection{Checking out a monster sighting}

In 2003 a team of radio astronomers reported evidence for the
discovery of a supermassive black hole binary in the bright radio
galaxy 3C~66B~\cite{Sudou}, which is located about 90~Mpc from Earth.
Their claim was based on very long baseline interferometry (VLBI)
observations of the galaxy, in which the radio core of the galaxy was
seen to move slightly over the course of 15 months in a manner
consistent with an elliptical orbit with a period of 1.05 year.  This
suggested the presence of a binary system with a total mass near $5
\times 10^{10}\,M_\odot$.  Remarkably, such a system would also be
expected to merge in $\sim$ 5 years due to energy loss by GW emission.

Jenet, Lommen, Larson \& Wen determined that pulsar timing could be used
to check this claim, since the gravitational waves from the binary
would cause pulse time-of-arrival variations of up to several
microseconds~\cite{Jenet3C66B}.  They analyzed seven years of archival
Arecibo timing data for PSR 1855+09 and found no such variation,
definitively ruling out the proposed binary system in 3C~66B.

\subsection{Catching a merger anywhere in the sky}

Binary neutron star systems are benchmark sources for ground-based GW
detectors because binary pulsars give a glimpse of the population
(see discussion in
\cite{KalogeraBNS}) and efficient GW emission during the inspiral
phase just before merging makes them detectable out to considerable
distances, currently tens of megaparsecs.  Black-hole-and-neutron-star
(BHNS) and binary black hole (BBH) systems with stellar-mass black holes can be
detected at even greater distances; population synthesis studies
suggest that the net detection rates for those sources are likely
to be
comparable~\cite{OSNSBS,KalogeraBBH}.  Because GW detectors have wide
antenna patterns, signals from these events can be detected from
essentially anywhere in the sky and at any time.

The most sensitive search published to date for these sources used
LIGO S5 data and templates for binary inspirals with total mass up to
$35\,M_\odot$~\cite{S5CBClow}.
No significant signal candidate was detected.  The
LSC interpreted this null result using a population model based on
the assumption that the rate of mergers
in each nearby galaxy is proportional to its blue light luminosity, as a
tracer of massive star formation.  The upper limits obtained from
a total of 18 calendar months of LIGO data, in
units of merger rate per year per $L_{10}$ (defined as $10^{10}$ times
the blue light luminosity of the Sun), were $0.014$, $0.0036$ and
$0.00073$ for binary neutron star, BHNS, and
BBH systems, respectively, calculated assuming black
hole masses of $5 \pm 1\,M_\odot$.  These limits are still far
from the theoretically expected rates, but are motivating the
numerical relativity community to improve waveform calculations
and the data analysis community to find better ways to search
for inspirals with higher masses,
significant spin and/or high mass ratio~\cite{EOBNR,EOBresum}.

\subsection{Being vigilant for arbitrary bursts}

Supernova core collapse and several other plausible signals are not
modeled well enough to use matched filtering,
either because the astrophysics is
not completely known or because the physical parameter space is too
large to effectively cover with a template bank.  Even for the
important case of BBH mergers, which numerical
relativity calculations are now having considerable success in
modeling~\cite{NRadvances}, the physical parameter
space allows for a wide variety of waveforms.  Thus it is important to
search for arbitrary transient signals (bursts) in the GW data, and
data analysis methods have been implemented which robustly detect a
wide range of signals without advance knowledge of the waveform.

\begin{figure}[bt]
\begin{center}
\includegraphics[width=11cm]{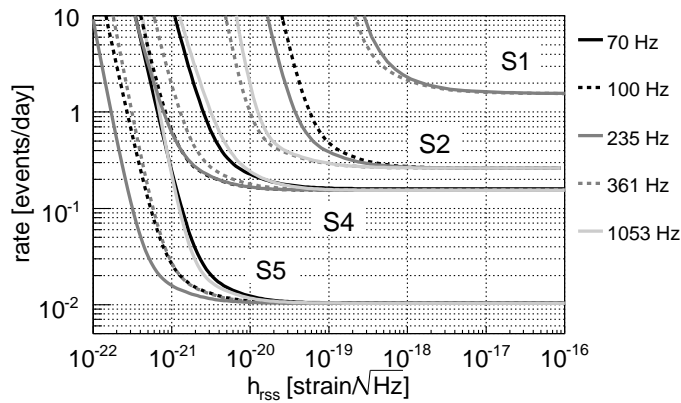}
\caption{Burst rate versus strength limit plot for sine-Gaussian
  waveforms~\cite{S5y1Burst}.  The
  area above each curve is excluded at 90\% confidence level.  Results
  are shown from the LIGO S1, S2 and S4 science runs in addition to 
  the first calendar year of the S5 run (labeled ``S5'' here), showing
  the progression of increasing sensitivity (to the left) and
  observation time (downward).
  Reprinted figure 8a with permission from B.\ P.\ Abbott {\it et al},
  Physical Review D {\bf 80}, 102001 (2009).  Copyright (2009) by the
  American Physical Society.}
\label{fig:BurstPlotSG}
\end{center}
\end{figure}

So far the IGEC network of bar detectors is the observation time
champion, having collected enough data since 1997 to establish an
upper limit of $1.5$ per year on the rate of strong GW
bursts~\cite{IGECburst}, with looser rate limits on weaker bursts.
More recent IGEC data extended their sensitivity to somewhat weaker
bursts~\cite{IGEC2burst}, but with looser rate limits of $\sim$8.5
per year.  On the other hand, LIGO is the sensitivity champion.  The
latest published burst search results~\cite{S5y1Burst},
from the first calendar year of
the LIGO S5 run, set upper limits on the rate of bursts arriving at
Earth as a function of signal waveform and amplitude, expressed as the
root-sum-squared GW strain calculated from both polarization
component amplitudes at the Earth:
\begin{equation}
h_{\textrm{rss}} = \sqrt{{\int_{-\infty}^{+\infty} \left( |h_+(t)|^2 +
  |h_{\times}(t)|^2 \right) \, dt}} \,\,.
\end{equation}
Figure~\ref{fig:BurstPlotSG}
shows the limits set by burst searches in the S5 run (first calendar
year only) and earlier science runs for ``sine-Gaussian'' waveforms
with $Q=9$ and central frequencies up to $\sim$1~kHz.
The area above each curve is excluded at 90\% confidence level, {\it
  i.e.}\ the curve traces out the 90\% upper limit on the rate for a
given waveform assuming a hypothetical population with fixed
$h_{\textrm{rss}}$.  Sufficiently loud bursts of any form have the
same rate limit, $3.75$ per year, as the efficiency of the analysis pipeline
approaches unity.

The signal strength can also be related in a robust way to GW energy
emission from a source at a known or assumed distance $r$.
For instance,
the S5 first-year search mentioned above would have
been sensitive to an event in the Virgo galaxy cluster ($r \simeq
16$~Mpc) that emitted $\sim 0.25\,M_\odot c^2$ of GW energy in a burst
with a dominant frequency of $\sim 150$~Hz \cite{S5y1Burst}.  These
searches constrain populations of sources such as binary mergers of
intermediate-mass black holes, although so far only preliminary
quantitative studies have been made with realistic simulated
waveforms~\cite{ninja}.

\subsection{Tuning in to spacetime shivering}

The Big Bang may have left behind a stochastic background of
gravitational waves, isotropic like the cosmic microwave background
(CMB) but carrying information about much earlier fundamental
processes in the early universe; see \cite{Maggiore} for a review and
references in \cite{S5stoch} for updates on the details of plausible
processes.  A stochastic background, isotropic or not, can also be
produced by a large number of overlapping astrophysics sources such as
binary mergers, cosmic (super)strings, or core-collapse supernovae.
The GW signal generally has the form of random ``noise'' with a
characteristic power spectrum, though it can be distinguished from
true instrumental noise by testing for a common signal in multiple
detectors for which the instrumental noise is known to be uncorrelated.

Jenet {\it et al} have used pulsar timing to search for low-frequency
stochastic gravitational waves in archival and newly-obtained data for
seven pulsars spanning intervals from $2.2$ to $20$
years~\cite{PTAstoch}.  They detected no signal but placed limits on
the GW energy density assuming different power-law distributions as a
function of frequency.  From these they also derive limits on mergers
of supermassive binary black hole systems at high redshift, relic
gravitational waves amplified during the inflationary
era~\cite{Grishchuk2005}, and a possible population of cosmic
(super)strings~\cite{DamourVilenkin}.

The LSC and Virgo have used LIGO data to search for a stochastic GW
signal in the vicinity of 100~Hz by measuring
correlations in the data from the Hanford and Livingston
interferometers to test for a signal well below the noise level of
either instrument.  A recently published paper~\cite{S5stoch} used the
data from the full S5 science run to set a limit on the energy density
in GWs as a fraction of the critical energy density needed to close
the universe.  The result, assuming a frequency-independent spectrum,
was $\Omega_0 < 6.9 \times 10^{-6}$ at 95\% confidence.  This direct
limit surpasses the indirect limits from Big Bang nucleosynthesis and
the CMB and constrains early-universe models.  It also imposes
constraints on a possible population of cosmic strings in a different
part of the parameter space than the pulsar timing result does.

\subsection{Summary: impact of null results}

From this sampling of search results, most published in the past few
years, one can see the beginnings of rich astrophysics coming out of
GW observations.  The searches are now placing meaningful constraints
on some individual objects and events, source populations (either real
or speculated), and the total energy density of gravitational waves in
the universe.  Many more analyses are in progress, and null results
will surely continue to provide interesting information.

\section{The (future) impact of detected signals}

As I write this sentence, I can see\footnote{I checked the web
  page
  http://www.ldas-sw.ligo.caltech.edu/ligotools/runtools/gwistat/,
  which reports the current status of operating GW detectors, on 29
  November 2009.} that
the two LIGO 4-km detectors and Virgo are all collecting science-mode
data at this particular moment (as part of the ongoing S6/VSR2 science
run), while AURIGA, EXPLORER and NAUTILUS are also collecting good
data.
GEO\,600 is being upgraded to ``GEO-HF'' with a focus on improving the
sensitivity for frequencies above $\sim$400~Hz~\cite{GEO-HF} and
will collect more data over the next several years.
It is possible that the first unambiguous GW signal is in the
data already collected but not yet fully analyzed, or will soon be
recorded.

Besides proving without a doubt that GWs exist and can be detected,
even a single detection would give us invaluable information about the
source from the waveform properties.  In the case of a binary
inspiral, the ``chirp'' rate and possible modulation reflect the
component object masses and spins; for the ringdown of a perturbed
black hole, the damped-sinusoid frequency and decay rate reveal the
mass and spin; for a spinning neutron star, the signal amplitude and
polarization content indicate its ellipticity and spin axis
inclination; and so on.  A signal that does not match any of the
standard models could confirm a speculative source type or reveal an
unanticipated one.  The reconstructed sky position of the source may
point to a galaxy and thus pin down the distance.  If the signal is
associated with an astronomical event or object observed by other
means---such as a GRB, optical or radio afterglow, supernova, neutrino
detection, or known pulsar---then the complementary information will
provide a clearer view of the nature of the source and emission
mechanisms.

Preparations are well underway for major upgrades to the ground-based
GW detector network in the form of Advanced LIGO~\cite{AdLIGO} and
Advanced Virgo~\cite{AdVirgo}, which will have an order of magnitude
better sensitivity than the current instruments.  When these detectors
begin operating in 2014 or 2015, we can expect regular detections of
binary inspirals
and excellent prospects for
detecting various other signals.  The detection of multiple signals of
the same type will enable population surveys that reveal the origin
and evolution of such sources.  Proposals for other large
interferometric detectors, in particular LCGT in Japan~\cite{LCGT} and
AIGO in Australia~\cite{AIGO}, would add significantly to the
capabilities of the current detector network.  Prospects are good for
detecting low-frequency signals with pulsar timing arrays on about the
same time scale~\cite{PulsarTiming}, while the space-based detectors
are due to be launched some years later to open up the intermediate
frequency band.
Conceptual designs for ``third-generation'' ground-based detectors
such as the Einstein Telescope (ET)~\cite{ETweb,ETpaper} are now being
proposed with the goal of improving over the sensitivities of the
``Advanced'' detectors by another order of magnitude.

\begin{figure}[bt]
\begin{center}
\includegraphics[width=6cm]{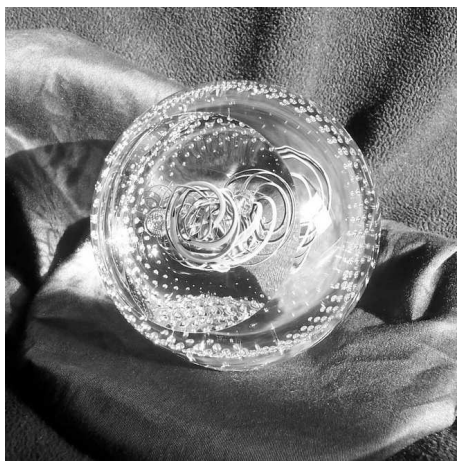}
\caption{Crystal ball for predicting the results of future 
   gravitational-wave searches.
   (Glass art by Henry Summa; photo by J G Shawhan.)}
\label{fig:CrystalBall}
\end{center}
\end{figure}

A comprehensive discussion of all the possible astrophysics that can
be addressed is beyond the scope of this article (but see the Living Review
by Sathyaprakash and Schutz~\cite{SathyaSchutz}, for example).
Instead, to illustrate some of the
key issues, let us look into a crystal ball (see
figure~\ref{fig:CrystalBall})
and make some predictions for what the future {\em might} hold.

\subsection{A wild guess at the future}

Near the end of the S6/VSR2 science run, LIGO and Virgo will record a
fairly significant inspiral event candidate, with a strength
corresponding to a false alarm rate of 1 per 160 years.  The
best-match template will have masses of $8.2\,M_\odot$ and
$1.46\,M_\odot$, representing a black hole and a neutron star.  The
all-sky burst search will also identify this as a candidate, but not
the strongest one in that search.  The reconstructed sky position will
be consistent with three galaxies within 50~Mpc.  Prompt follow-up
imaging with a robotic telescope will capture a weak, fading optical
transient in an elliptical galaxy within the favored sky region.
After careful internal review and debate, the LIGO and Virgo
collaborations will publish the complete diagnosis of the candidate,
calling it ``cautious evidence for a gravitational-wave signal''.

By 2015, pulsar timing array analyses will have produced greatly
improved upper limits on supermassive black hole binary mergers and
stochastic background processes, but no detections yet.

In the Spring of 2015, Advanced LIGO and Advanced Virgo will have been mostly
commissioned and (I speculate) will begin an 8-month science run while
still a factor of $\sim$2 away from their design sensitivities.  The
GEO-HF detector may continue to operate for part of that period.  The run
will yield two black hole--neutron star inspiral candidates, with an
expected background of $0.02$, and also two binary neutron star
inspiral candidates, with a background of $0.03$.  One of the binary
neutron star candidates will have a clear radio afterglow in prompt
follow-up observations with radio telescopes.  These event candidates
will be published together as the first clear detection of
gravitational waves.  Analysis of these events will also place strong
limits on ``extra'' GW polarization states beyond the two predicted by
general relativity.

All-sky searches for burst and periodic GW signals using the same data
will yield candidates that look promising but are not significant
enough to claim as detections.
Greatly improved upper limits will be published on GW
emission from known pulsars and on a stochastic background.

After further commissioning, Advanced LIGO and Advanced Virgo will
resume running at full sensitivity, joined the following year
by LCGT and AIGO.  I imagine that analysis of two years of data
will yield the following:
\begin{itemize}
\item 15 binary neutron star candidates with a background of $2.2$.
  Two of the candidates will correspond to short-hard GRBs, one of
  which is localized in a galaxy with a measured redshift of $0.07$.
  Based on this information, the emitted GW energy will turn out to be
  consistent with the theoretical prediction.
\item 18 black hole--neutron star candidates with a background of
  $4.1$.  Two of these will correspond to GRBs, one also with a
  high-energy neutrino.
\item Comparison of GW inspiral times with GRB times will confirm that GWs
  travel at the speed of light.
\item 6 binary black hole candidates with a background of $1.7$.  The
  masses and spins of the candidates will be inferred, giving a
  preliminary look at their distributions.
\item 4 burst candidates with a background of 0.15.  One of
  them, with central frequency 310~Hz, will correspond to a weak long GRB
  with no measured redshift.
\item A periodic GW signal will be detected from the Crab pulsar,
  corresponding to $0.12$\% of the total spin-down energy.  This
  result will be used to constrain models of neutron star formation
  in supernovae.
\item Periodic GW signals will also be detected from Sco X-1 (using
  data collected during a 3-month period with the detectors in a
  narrow-band configuration) and from 5 unseen neutron stars.
\item Stochastic GW searches will detect signals from two low-mass
  X-ray binaries (LMXBs) besides Sco X-1, and will place much stricter
  limits on cosmic string models.
\end{itemize}

Around the same time, pulsar timing analyses will detect GWs from
supermassive black hole binary systems in two galaxies, and will rule
out another large area of the parameter space for cosmic string models.
Gravitational-wave astronomy will thus be in full swing by the time that
LISA and DECIGO are launched and open up new frequencies for GW
observations.

\subsection{Notes on this exercise}

It may have been indulgent to speculate so specifically about what the
future may bring, and the details are obviously fictional.
However, I think the scenario above is actually fairly conservative
and illustrates several of the scientific findings that can be derived
from the observations.  It also reflects many of the issues the
GW community will have to deal with, such as
borderline-significant event candidates, samples of event candidates
with non-negligible backgrounds, and the role of information from
electromagnetic observations.

\section{The impact of detections on the field of gravitational-wave science}

The transition envisioned above, from always setting upper limits to
being able to claim some detections, will call for some changes in
strategy to make optimal use of the detectors and of the data for
science results.  In this section we discuss a few such areas.

\subsection{Tuning choices for advanced detectors}

Interferometric detectors may be operated in different ways in order
to optimize the noise characteristics according to scientific
priorities.  For instance, the Advanced LIGO and Advanced Virgo
detectors are designed to be limited by quantum noise at low and high
frequencies; by reducing the laser intensity, one can reduce the
radiation pressure noise at low frequency at the cost of increasing
the shot noise at high frequency.  Interferometer configurations with
signal recycling, such as Advanced LIGO and Advanced Virgo, allow
additional tuning options through changing the reflectivity and
(microscopic) detuning phase shift of the signal recycling mirror.
Optimal tunings have been considered for individual signal types as
well as some combinations~\cite{AdLIGOtuning}.  Of course, an
interferometer can only operate in one mode at a time.  Detection of
one or more GW signals may motivate re-tuning the interferometer to
focus on a certain class of signals, either temporarily or for the
rest of the run.  It may also be useful to tune different
interferometers differently, {\it e.g.}\ one of the Advanced LIGO Hanford
interferometers could be optimized for low frequency while the other
is optimized for higher frequency.

\subsection{Support for additional detectors}

The first direct detection of a GW signal will erase any lingering
doubts about whether GW detectors really work, and will bring a new
focus to the science which can be done with GW observations.
That should make a
stronger case for additional detectors on the ground
({\it e.g.}\ LCGT, AIGO, ET) as well as boosting support for 
detectors in space (LISA, DECIGO).
Furthermore, the designs of new detectors may be
influenced by the view of what measurements are most important 
based on what has been detected so far.

\subsection{Changes in philosophy}

Currently there is a big emphasis in the GW detection community on
achieving near-perfect certainty in the first GW signal detection.
Typically this results in raising the signal strength threshold so
that the false alarm rate is extremely low, but that also reduces the
sensitivity of the search.  However, having certified one or more
events as genuine increases our belief in other candidates of the same
type.  Thus we can choose to relax the signal strength threshold to
include more candidates in a search, even if doing so also includes
more background---the benefit from having more real signals in the
sample to study may outweigh the negative effects of the additional
background.

\section{Summary}

Many of the gravitational-wave searches that have been performed in
recent years have provided useful astrophysical information despite
yielding no confirmed GW signal candidates.  Thus, one can say that we
are already doing gravitational-wave astronomy.  Actual detections,
when they finally start coming, will enable us to address a much wider range
of astrophysics questions.
And here is what will be exciting: {\em Everything}.

\ack

I would like to thank the organizers of the Eighth Edoardo Amaldi
Conference on Gravitational Waves for giving me this opportunity to
review and interpret the current state of gravitational-wave
astronomy.  Of course, my colleagues in the gravitational-wave
community are responsible for the observational results themselves,
and my views of the astrophysics and of the field have been shaped by
discussions with many of them---too many to thank individually.
I was particularly inspired by a 2007 seminar by Ben Owen
entitled ``Why LIGO results are already interesting''.
I gratefully acknowledge the support of the National Science
Foundation through grant PHY-0757957.

This article has been assigned LIGO document number P0900289-v4.

\end{document}